\documentclass[3p]{elsarticle}

\usepackage{graphics}
\usepackage{epsfig}

\usepackage{amssymb}

\begin{document}

\begin{frontmatter}


\title{Backward-angle photoproduction of $\omega$ and $\eta'$ mesons from protons at $E_{\gamma}=1.5-3.0$~GeV }

\author[a,b]{ Y.~Morino\corref{cor1}} 
\cortext[cor1]{Corresponding author}
\ead{yuhei.morino@riken.jp}
\author[c]{ Y.~Nakatsugawa}
\author[a]{ M.~Yosoi} 
\author[e]{ M.~Niiyama} 
\author[b]{ D.~S.~Ahn} 
\author[f]{ J.~K.~Ahn} 
\author[a]{ S.~Ajimura}
\author[g]{ W.~C.~Chang}
\author[a]{ J.~Y.~Chen} 
\author[h]{ S.~Dat${\rm \acute{e}}$} 
\author[i]{ H.~Fujimura} 
\author[j]{ S.~Fukui} 
\author[k]{ K.~Hicks} 
\author[a]{ T.~Hiraiwa} 
\author[a]{ T.~Hotta} 
\author[l]{ S.~.H.~Hwang} 
\author[l]{ K.~Imai} 
\author[m]{ T.~Ishikawa} 
\author[j]{ Y.~Kato}  
\author[n]{ H.~Kawai} 
\author[o]{ M.~J.~Kim} 
\author[a]{ H.~Kohri}
\author[a]{ Y.~Kon}
\author[g]{ P.~J.~Lin}
\author[n]{ K.~Mase} 
\author[p]{ Y.~Maeda}
\author[m]{ M.~Miyabe}  
\author[m]{ N.~Muramatsu} 
\author[a]{ T.~Nakano} 
\author[a]{ H.~Noumi} 
\author[l]{ Y.~Ohashi}
\author[a]{ T.~Ohta} 
\author[a]{ M.~Oka} 
\author[e]{ J.~D.~Parker}
\author[q]{ C.~Rangacharyulus} 
\author[a]{ S.~Y.~Ryu}
\author[n]{ T.~Saito} 
\author[a]{ T.~Sawada}
\author[m]{ H.~Shimizu}  
\author[r]{ E.~A.~Strokovsky} 
\author[s]{ Y.~Sugaya} 
\author[t]{ M.~Sumihama}
\author[s]{ K.~Suzuki} 
\author[o]{ K.~Tanida} 
\author[a]{ A.~Tokiyasu}
\author[n]{ T.~Tomioka}
\author[a]{ T.~Tsunemi}
\author[u]{ M.~Uchida} 
\author[a]{ R.~Yamamura}
\author[a]{ T.~Yorita}
\address[a]{Research Center for Nuclear Physics, Osaka University, Ibaraki, Osaka 567-0047, Japan} 
\address[b]{RIKEN Nishina Center for Accelerator-Based Science, Wako, Saitama 351-0198, Japan}
\address[c]{KEK, High Energy Accelerator Research Organization, Tsukuba, Ibaraki 305-0801, Japan}
\address[e]{Department of Physics, Kyoto University, Kyoto 606-8502, Japan}
\address[f]{Department of Physics, Pusan National University, Busan 609-735, Korea}
\address[g]{Institute of Physics, Academia Sinica, Taipei 11529, Taiwan}
\address[h]{Japan Synchrotron Radiation Research Institute, Sayo, Hyogo 679-5143, Japan} 
\address[i]{Wakayama Medical University , Wakayama 641-8509, Japan}
\address[j]{Department of Physics and Astrophysics, Nagoya University, Nagoya, Aichi 464-8602, Japan}
\address[k]{Department of Physics and Astronomy, Ohio University, Athens, OH 45701, USA} 
\address[l]{Japan Atomic Energy Agency, Tokai-mura, Ibaraki 319-1195, Japan}
\address[m]{Research Center for Electron Photon Science, Tohoku University, Sendai, Miyagi 982-0826, Japan} 
\address[n]{Department of Physics, Chiba University, Chiba 263-8522, Japan}
\address[o]{Department of Physics and Astronomy, Seoul National University, Seoul 151-747, Korea}
\address[p]{Proton Therapy Center, Fukui Prefectural Hospital, Fukui 910-8526, Japan}
\address[q]{Department of Physics and Engineering Physics, University of Saskatchewan, Saskatoon SK S7N 5E2, Canada}
\address[r]{Joint Institute for Nuclear Research, Laboratory of High Energy Physics, 141980, Dubna,Russia}
\address[s]{Department of Physics, Osaka University, Toyonaka, Osaka 560-0043, Japan}
\address[t]{Gifu University, Gifu 501-1193,Japan} 
\address[u]{Department of Physics, Tokyo Institute of Technology, Tokyo 152-8551, Japan}

\begin{abstract}
We report the measurement of differential cross sections for $\omega$ and $\eta'$ photoproduction from protons 
at backward angles~($-1.0<\cos\Theta_{C.M}^{X}<-0.8$) using linearly polarized photons at $E_{\gamma}=$$1.5-3.0$~GeV.
Differential cross sections for $\omega$ mesons are larger than the predicted $u$-channel contribution 
in the energy range $2.0\leq\sqrt{s}\leq2.4$~GeV.
The differential cross sections for $\omega$ and $\eta'$ mesons become closer to the predicted $u$-channel contribution at $\sqrt{s}>2.4$~GeV.
A bump structure in the $\sqrt{s}$ dependence of the differential cross sections for $\eta'$ mesons was observed at $\sqrt{s}\sim$2.35~GeV.
\end{abstract}


\end{frontmatter}
Measurements of meson photoproduction provide a good tool to study nucleon resonances.
Many nucleon resonances have been identified from experimental and theoretical study of
$\pi$N scattering and $\pi$ photoproduction.
It is well known that a large number of resonances predicted by the constituent 
quark model remain to be discovered~(missing resonance problem)\cite{bib:pdg,bib:mis}.
Some of the missing resonances may not be observed due to the weak coupling to the pion,
but could be observed in the photo-production of other mesons.
Among various channels, $\eta$ and $\eta'$ photoproduction are of  special interest 
since these mesons possess a large component of $s\bar{s}$. 
The interpretation of the large branching ratio of S$_{11}$(1535)$\rightarrow p\eta$ decay has been 
a topic of a much discussion; it could be a dynamically-generated state or 
a conventional three quark state\cite{bib:e1,bib:e2,bib:e3}. 
A systematic study of nucleon resonances with large 
couplings to $\eta$ and $\eta'$ mesons will give important information to solve this controversy.
Recently, differential cross section and polarization variable of $\eta$, $\eta'$ and $\omega$ mesons 
have been measured in experiments like CB-ELSA, GRAAL and 
CLAS with large acceptance spectrometers\cite{bib:clas1,bib:gral1,bib:elsa1,bib:elsa2,bib:clas2,bib:gral2}.
Evidence and indication of new resonances have been obtained from partial wave analysis~(PWA)
of their results, although the list of resonances depends on models. 
There is a significant contribution from nucleon resonances in the differential cross section
of meson photoproduction at large scattering angle~($\Theta_{C.M}^{X}\sim\pi/2$) 
at $\sqrt{s}\sim2$~GeV.

At backward angles in the center of mass system, it is expected that the contribution from $u$-channel exchange of Regge trajectories 
becomes significant. 
The differential cross section from the $u$-channel baryon exchange is expected to behave following a power law of $s$.
In general, the differential cross section from the $u$-channel is much smaller than the one from the $t$-channel meson exchange.
On the other hand, the angular distribution of mesons from nucleon resonances could have a rapid change at 
forward and backward angles when the nucleon resonances have high angular momenta.
The contribution of nucleon resonances with high angular momenta tends to be stronger at forward and backward angles 
than at intermediate angles.
Therefore, the differential cross section at backward angles is sensitive and a good tool to identify and search for
nucleon resonances with high angular momenta.
A bump structure in the $s$ dependence of differential cross section
at very backward angles has been observed at SPring-8/LEPS\cite{bib:lep2}.
A new measurement was carried out at SPring-8/LEPS with a time projection chamber~(TPC) surrounding the target
in order to detect decay products of hadrons.
Production of $\omega$ and  $\eta'$ mesons was clearly identified  by detecting protons at the  LEPS forward spectrometer 
and pions at the TPC.
In comparison with the previous LEPS experiment, background events of $\omega$ and  $\eta'$ signals 
were reduced substantially by using TPC \cite{bib:lep2}.
In addition, $E_{\gamma}$ was extended to 3.0~GeV.
In this article, we report the differential cross sections of $\omega$ and  $\eta'$ photoproduction at 
backward angles~($-1.0<\cos\Theta_{C.M}^{X}<-0.8$) from protons in the energy range $E_{\gamma}$ = 1.5-3.0~GeV.

The experiment was carried out at the SPring-8/LEPS facility\cite{bib:lep3}.
A linearly polarized photon beam in the energy range from 1.5 to 3.0 GeV was produced by backward-Compton scattering~(BCS) 
from the head-on collision between laser photons and 8-GeV electrons in the storage ring. 
Both 355-nm and deep-UV 257-nm lasers were used to produce Compton-scattered photons 
in the range of 1.5 to 2.4~GeV and 1.5 to 3.0~GeV, respectively.
The energy of a scattered photon was determined by measuring the recoil electron from
Compton scattering by a tagging counter. 
The energy resolution for the photon beam was about 15~MeV.
We used a liquid hydrogen~(LH$_{2}$) target with a length of 15~cm and a diameter of 40 mm.
The data was accumulated with $0.6\times10^{12}$ photons from 1.5 to 2.4~GeV at 
the target with the 355-nm laser, and with  $0.4\times10^{12}$ photons from 1.5 to 3.0~GeV with the 257-nm laser, respectively\cite{bib:laser}.
Half of the data with the 355-nm laser~(1.5-2.4~GeV) was taken with vertically polarized photons and 
the other half with horizontally polarized photons. 
The data with the 257-nm laser~(1.5-3.0~GeV) was taken only with vertically polarized photons.

The LEPS forward spectrometer consisted of a
dipole magnet, four multiwire drift chambers, a start counter (SC) 
just downstream of the target, a silica-aerogel ${\rm \check{C}}$erenkov counter (AC),  
and a time-of-flight (TOF) hodoscope placed downstream of the tracking detectors. 
In this measurement, one multiwire drift chamber was used instead of a silicon-strip vertex detector, 
which was a different setup from previous LEPS experiments.
The angular coverage relative to the photon beam of the forward spectrometer 
was about $\pm$0.25~rad  and $\pm$0.12~rad in the horizontal and vertical directions in the laboratory system, respectively.
A time projection chamber~(TPC) surrounding
the target was installed inside a solenoid magnet. 
The TPC was different from the one used in the previous LEPS experiment to install the LH$_{2}$ target\cite{bib:lep10}.
The strength of the magnetic field was 2~T.
The TPC had an active volume of hexagonal cylinder shape with a side length of 225 mm 
and a height of 750 mm.
The TPC volume was filled with P$_{10}$ gas (Ar:CH$_{4}$ 90\%:10\%).
The TPC had a hexagonal hole in the center with a side length of 69~ mm to install the target.
The azimuthal and polar angular coverages of the TPC were 2$\pi$ and 0.35-2.25~rad in the laboratory system, respectively.
The signals from the TPC were read through rectangular cathode pads with a length of 56~mm and 150~mm.
The typical spatial resolutions were 200-400~$\mu$m in the pad plane and
400-4000~$\mu$m in the beam direction depending on the direction of charged tracks.
Six scintillation counters surrounded the target inside the TPC and 
twelve scintillation counters were placed outside of the TPC.
The trigger in this experiment was generated from the coincidence among the tagging counter, 
any 1 of 6 inner counters and any 1 of 12 outer counters facing
a hit inner counter. 
This trigger required at least one charged particle
with $p_{\mathrm{T}}\geq0.09$~GeV/c in the TPC acceptance.
The trigger efficiency saturated at about 94\% in the geometrical acceptance of  
the scintillation counters.

\begin{figure}[thb]	
	\begin{center}	
	    \includegraphics[width=1\linewidth]{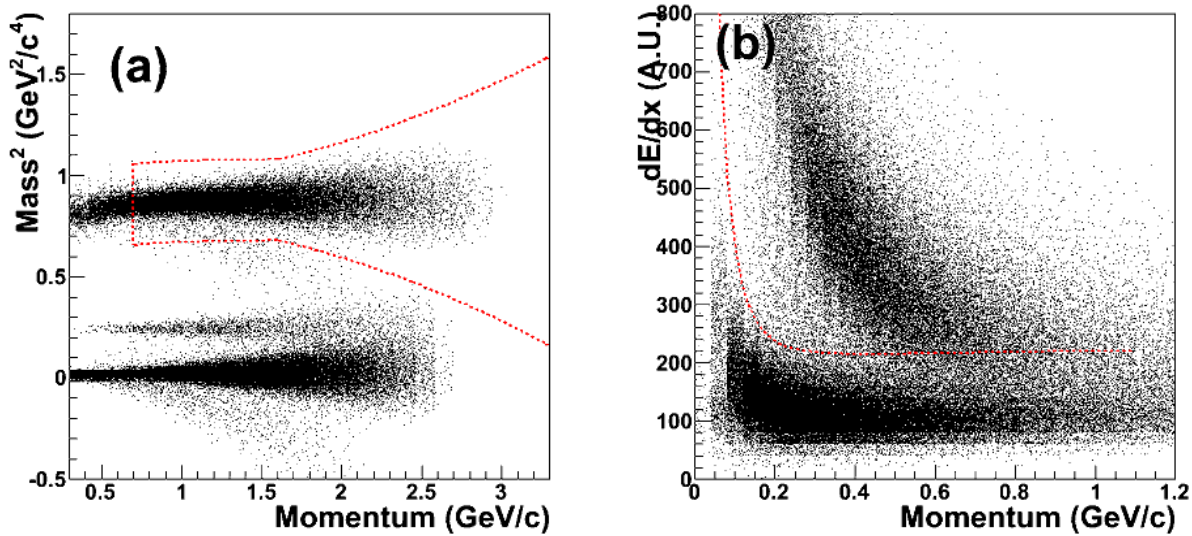}
	    \caption{(color online) (a) Correlation plot of mass squared and momentum of positive charged particles measured by the spectrometer. Dashed line shows boundary for proton identification. (b) Correlation plot of the energy deposition and momentum for positive charged particles measured by the TPC. Dashed line shows boundary for pion identification.}
	  \end{center}
\end{figure}
The $\omega$ and $\eta'$ mesons were measured via the following reactions.\\	
\begin{eqnarray}
  \gamma p & \rightarrow & p \omega  \rightarrow p \pi^+ \pi^- \pi^0  \\
  \gamma p & \rightarrow & p \eta'   \rightarrow p \pi^+ \pi^- \eta  
\end{eqnarray}
Protons were measured by the LEPS forward spectrometer.
Charged pions were detected in the TPC and $\pi^0$ or $\eta$ mesons were identified 
by missing mass information to select reaction (1) and (2).
It was required that the number of reconstructed charged tracks in the TPC was one or two.
The number of reconstructed electron tracks in the tagging counter was required to be one.
Protons were selected by the mass reconstructed from momentum and TOF information 
within 4$\sigma$ of the nominal value.
The momentum of a proton was required to be more than 0.7~GeV/c 
because $\omega$ and $\eta'$ mesons were produced only in this range.
Figure~1(a) shows the reconstructed mass square of positive charged particles 
as a function of momentum.
Admixture of tracks from particles misidentified as protons was estimated to be negligible~($<0.5$\%) and 
the momentum resolution for proton tracks was about 1\%.
Pions were identified from the mean of the energy deposition~($dE/dx$) in the TPC. 
Figure~1(b) shows measured $dE/dx$ of positive charged particles as a function of momentum.
The dashed line in Fig.~1(b) shows the boundary for pion identification with $>$98\% effeciency. 
The tracks with $dE/dX$ below the boundary line were identified as pions.
Although the $dE/dx$ cut for pion identification was not tight, tracks from particles misidentified as pions
were suppressed since a proton was already detected in the forward spectrometer.
The reconstructed tracks of pions had $p_{\mathrm{T}}$ larger than $\sim0.08$~GeV/c because of the center hole in the TPC.
The momentum resolution for tracks of the pions was 4\%-25\%, strongly depending on
momentum and polar angle of pions.
A reaction vertex point was reconstructed as the closest point between a track reconstructed 
in the spectrometer and a track in the TPC. 
The spatial resolution of the reaction vertex along the z direction was 2.6~mm.
Events produced from the target were selected by a cut on the z coordinate of the reaction vertex.
The effect of acceptance, efficiency and resolution of the spectrometer and the TPC 
were evaluated using a Monte-Carlo simulation with the GEANT3 code\cite{bib:geant}.

The systematic uncertainty for the target thickness including the target shape, 
fluctuations of the temperature, and pressure of the liquid hydrogen was estimated to be 2.0\%. 
The systematic error of the photon number normalization was estimated to be 3\% for data with the 355-nm laser and 
to be 4\% for data with the 257-nm laser, respectively. 
It includes fluctuation of proton yield per photon and transmission of the photon beam.
The systematic error of contamination from the events from the SC and the target cell was 1\%.
The systematic uncertainty for the efficiency of the spectrometer was 4\%, including
geometrical acceptance~(3\%), wire effeciency~(1\%), and proton identification effeciency~(2\%).
The  systematic uncertainty for the TPC efficiency was 4\%.

\begin{figure}[thb]	
	\begin{center}	
	    \includegraphics[width=1\linewidth]{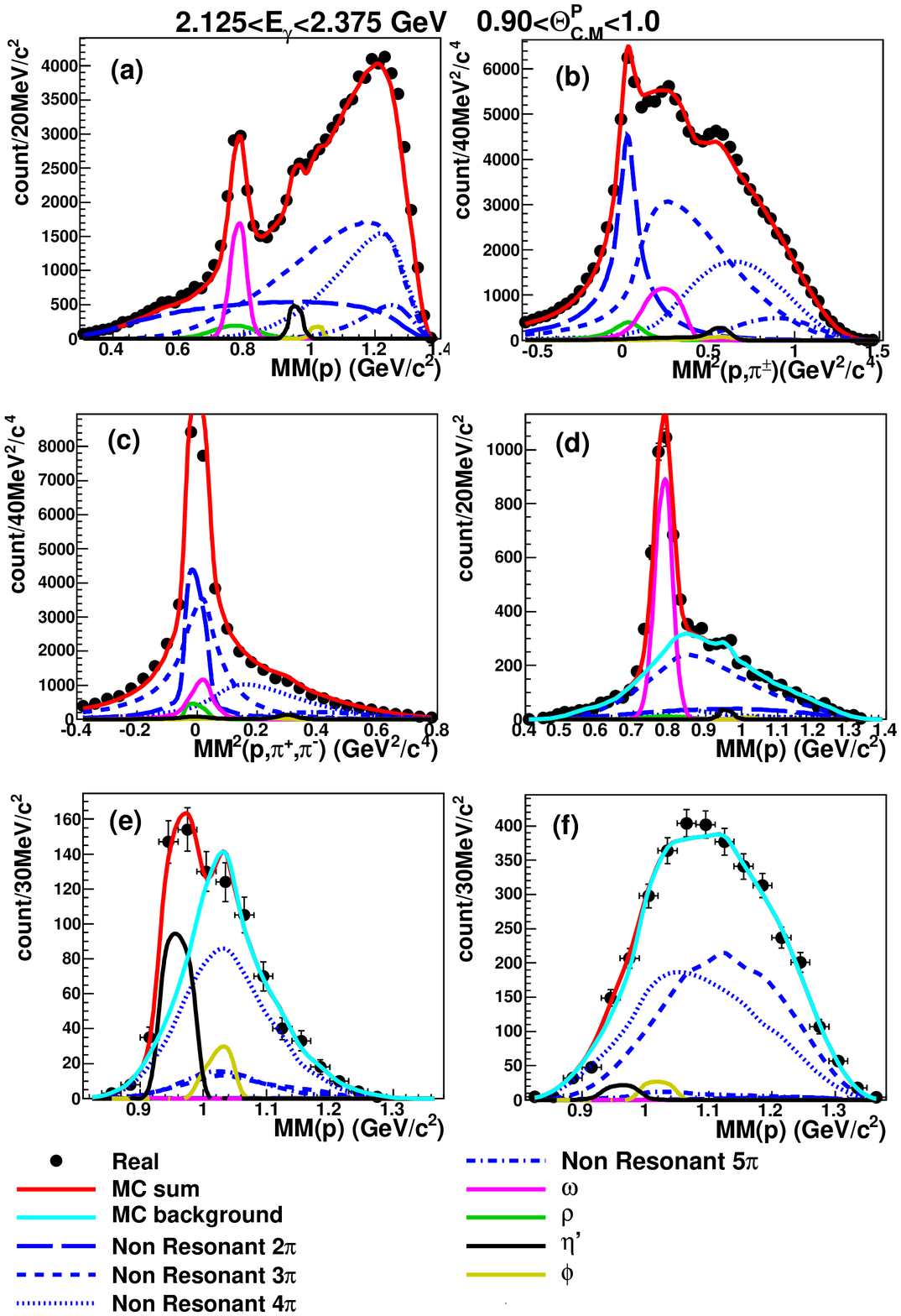}
	    \caption{(color online) Missing mass spectra for events with 
	    $2.125<E_{\gamma}<2.375$~GeV and $0.90<\cos\Theta_{C.M}^{P}<1.0$.
	    Circle points represent the experimental result.
	    Red solid lines represent sum of all contributions. 
	    Light blue lines represent sum of contributions without  
	    $\omega$ and $\eta'$ contribution for panel (d) and (e), respectively.
	    Blue dashed, dashed spaced, dotted and chain lines represent non-resonant 2, 3, 4, and 
	    5 $\pi$ production, respectively.    
	    Magenta, green, black and yellow lines represent $\omega$, $\rho$, $\eta'$
	    and $\phi$, respectively.
	    (a) $MM(p)$ (b) $MM^2(p,\pi^{\pm} X)$ (c) $MM^2(p,\pi^+, \pi^-)$ 
	    (d) $MM(p)$ with the $\omega$ selection cut (e) $MM(p)$ with the $\eta'$ selection cut.
	    (f) $MM(p)$ with the inverse $\eta'$ selection cut.}
	  \end{center}
\end{figure}
Figure~2(a) shows a missing mass spectrum for the $\gamma p \rightarrow p X$ reaction~($MM(p)$), 
where the photon energy is from 2.125 to 2.375~GeV and the scattering angle of protons is
$0.90<\cos\Theta_{C.M}^{P}<1.00$.  The spectra of missing mass squared
for the $\gamma p \rightarrow p \pi^{\pm} X$ and $\gamma p \rightarrow p \pi^+ \pi^- X$ reactions~($MM^2(p,\pi^{\pm})$ and 
$MM^2(p,\pi^+,\pi^-)$) are also shown in Fig.~2(b) and 2(c), respectively.  
The peaks of $\rho$/$\omega$ and $\eta'$ mesons are observed in Fig.~2(a) and the peaks of $\eta$ and $\pi^0$ mesons are not observed clearly
due to the trigger condition.  
Although the peak of $\omega$ mesons overlaps with the one of $\rho$ mesons,
$\omega$ mesons can be separated from $\rho$ meson by identification of the decay products.
In Fig.~2(b) and 2(c), the peaks due to
2$\pi$ and 3$\pi$ production are also seen.  
To reduce background events for $\omega$ and $\eta'$ production, selection cuts were
applied for $MM^2(p,\pi^{+/-})$ and $MM^2(p,\pi^+,\pi^-)$.  Figure~2(d)
and 2(e) show the $MM(p)$ distribution with the $\omega$ and the
$\eta'$ selection cuts, respectively.  
The $\omega$ selection cut was $-0.15<MM^2(p,\pi^+,\pi^-)<0.19$~GeV$^2$/c$^4$ 
and $0.05<MM^2(p,\pi^{+/-})<0.44$~GeV$^2$/c$^4$.
$0.24<MM^2(p,\pi^+,\pi^-)<0.36$~GeV$^2$/c$^4$ and $0.40<MM^2(p,\pi^{+/-})<0.72$~GeV$^2$/c$^4$ 
were applied for the $\eta'$ selection cut.
The background events for $\omega$ and $\eta'$ signals were reduced drastically by applying
the selection cuts.
Figure~2(f) shows the $MM(p)$ distribution with an inverse $\eta'$ selection cut.
The condition for $MM^2(p,\pi^+,\pi^-)$ was reversed for the inverse 
cut~(($MM^2(p,\pi^+,\pi^-)<0.24$~GeV$^2$/c$^4$ or $MM^2(p,\pi^+,\pi^-)>0.36$~GeV$^2$/c$^4$) 
and $0.40<MM^2(p,\pi^{+/-})<0.72$~GeV$^2$/c$^4$).
This plot was prepared for the demonstration of understanding of the background shape.

Yields of $\omega$ and $\eta'$ mesons were extracted via the $MM(p)$ distribution 
with the $\omega$ and $\eta'$ selection cut~(Fig.~2(d)and 2(e)).
Background events for $\omega$ and $\eta'$ signals consisted of several reactions.
To evaluate these contributions, missing mass distributions of all reactions were prepared by 
using Monte-Carlo simulation.
The relative yield of the each reaction was determined by minimizing of the $\chi^2$ 
between the superposition of the prepared distributions and the experimental data~(a template fit).
Events including non-resonant pions~(from 2 to 5) and one proton were generated in the free N-body space as 
background components.
Photoproductions of $\eta$, $\eta'$, $\rho$, $\omega$, and $\phi$ mesons were also generated  as known resonances.
The $MM(p)$, $MM^2(p,\pi^{\pm})$, and $MM^2(p,\pi^+,\pi^-)$ shapes of these components 
were obtained by using Monte-Carlo simulation to take into account the detector response.
The shape of $MM(p)$ with a loose $\omega$ selection cut~($0.1<MM^2(p,\pi^{\pm})<0.4$~GeV$^2$/c$^4$)
was also prepared to determine to the $\rho/\omega$ ratio.
The distributions of $MM(p)$, $MM^2(p,\pi^{\pm})$, $MM^2(p,\pi^+,\pi^-)$, and  $MM(p)$ with the loose $\omega$ selection cut
were used as the constrain of the template fit simultaneously.
Events were put in photon energy bins with an interval of 62.5~MeV and proton scattering angle bins
with a 0.05~$\cos\Theta_{C.M}^{P}$ interval.
The template fit was performed for each bin on the photon energy and on the scattering angle of proton.
The reduced $\chi^2$ was 0.9 at minimum and 2.7 at maximum, depending on the angular and energy bins. 
The contribution of each reaction in the fitting is shown in Fig.~2(a),~2(b), and 2(c),
where the red solid lines represent the sum of all contributions and should be compared 
with the experimental results.
The $MM(p)$ shapes with the $\omega$ and $\eta'$ selection cuts in each reaction were also obtained. 
To determine the background shape, these were summed up  according to the relative yields in the fitting result,
except the resonances corresponding to the selection cuts~($\omega$ or $\eta'$).
The normalization of the background was determined by template fits with the background and signal shapes
for the $MM(p)$ distributions with the selection cuts.
The obtained background and each component are shown in Fig.~2(d), 2(e), and 2(f).
In Fig.~2(d), 2(e), and 2(f) the light blue solid lines represent the determined background and 
the red solid lines represent the sum of all contributions.
The sum of all contributions reproduces the experimental results successfully, including the result with
the inverse cut~(Fig.~2(f)).
The yields of $\omega$ and $\eta'$ signals were extracted by the subtraction of the background. 
The yields were corrected by the efficiency evaluated by the Monte-Carlo simulation to determine differential 
cross sections.
The typical efficiency of $\omega$ was 23\% at $-1.00<\cos\Theta_{C.M}^{\omega}<-0.95$
and 9\% at $-0.85<\cos\Theta_{C.M}^{\omega}<-0.80$, respectively.
The typical efficiency of $\eta'$ was 6\% at $-1.00<\cos\Theta_{C.M}^{\omega}<-0.80$.

The systematic uncertainty for the trigger efficiency was 5\% including 
the uncertainty of the efficiency in the acceptance~(3\%) and the uncertainty caused by the acceptance of  
the trigger counters~(4\%).
The systematic uncertainty for the efficiency for the $\omega$ and $\eta'$ selection cut was 
estimated by applying loose selection cut and by varying the cut boundary.
It was determined to be 3\%.
The systematic uncertainty for the background shapes was estimated by varying 
the fit conditions~(without the 5$\pi$ reaction, reducing the fit constrain, 
strategy of minimizing $\chi^2$, and the combination of these). 
It was determined to be 3\% and 5\% for $\omega$ and $\eta'$, respectively.
Since there was only the vertical polarization data in $E_{\gamma} = 2.4-3.0$~GeV,
the effect of polarization on the efficiency was included in the systematic uncertainty for this energy range. 
The effect on the efficiency was  evaluated at the given beam asymmetry by the Monte-Carlo simulation.
The beam asymmetry in $E_{\gamma} = 2.4-3.0$~GeV was assumed be less than 0.1 
from the measured beam asymmetry in $E_{\gamma} = 1.5-2.4$~GeV.
The systematic error for the effect of polarization was 2-7\% depending on the scattering angles.

\begin{figure}[thb]	
	\begin{center}	
	    \includegraphics[width=1\linewidth]{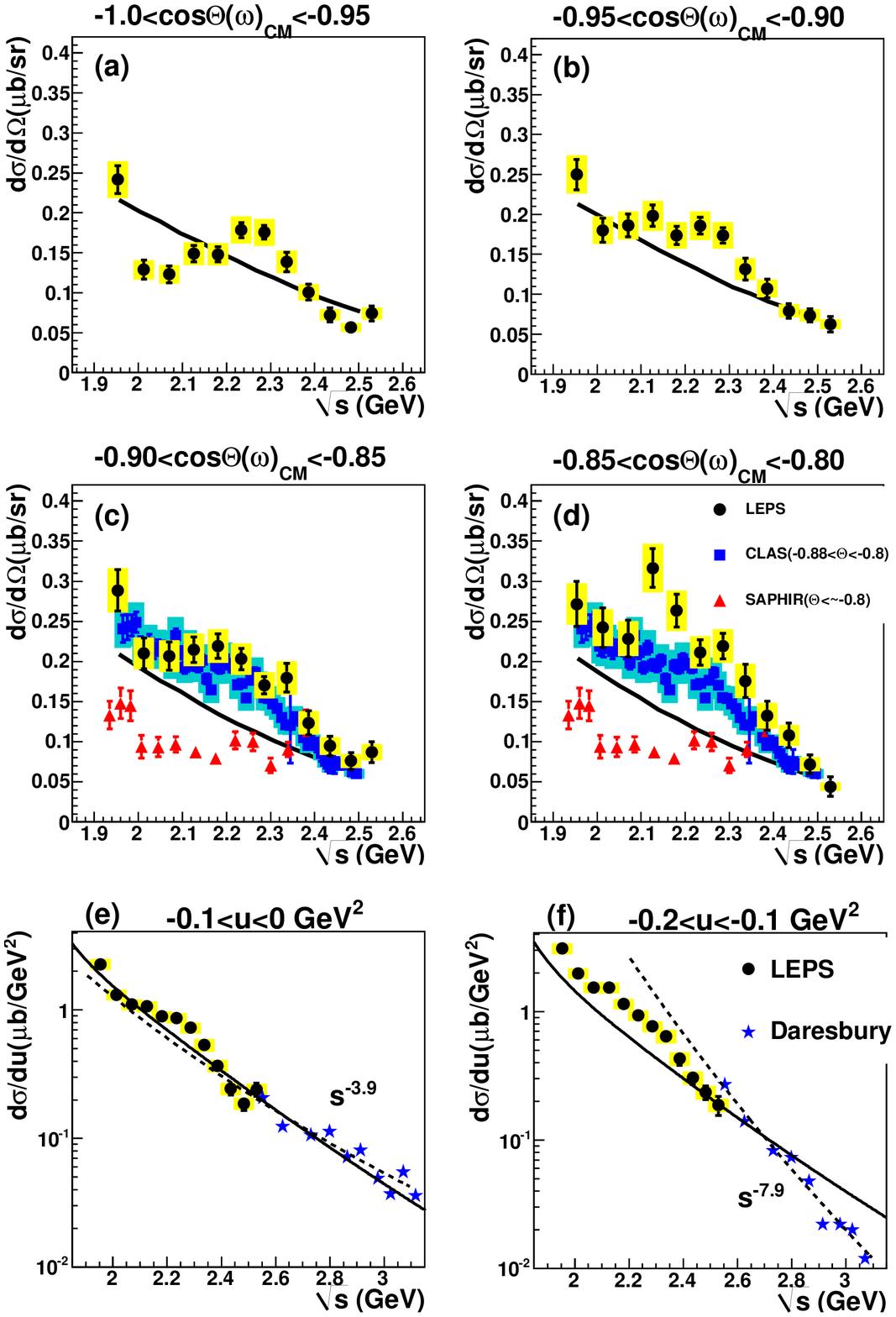}
	    \caption{(color online) Differential cross sections for $\omega$ photoproduction as 
	    a function of $\sqrt{s}$. 
	    The black circles are the present results.
	    Shaded bars represent systematic uncertainty.
	    The red triangles, blue squares and blue stars are the experimental results from 
	    SAPHIR, CLAS and Daresbury, respectively\cite{bib:clas1,bib:sap1,bib:dar1}.
	    Smooth lines represent theoretical calculation and 
	    dashed lines represent the result of a power law fit for the Daresbury results\cite{bib:alex1}.}
	  \end{center}
\end{figure}
Figure~3 shows the differential cross sections for $\omega$ production as a function of $\sqrt{s}$. 
Each panel shows the result at the $\omega$ scattering angle bin~((a)$\sim$(d) ) and the $u$ interval bin~((e) and (f)).
In $-0.9<\cos\Theta_{C.M}^{\omega}<-0.8$,  the results of CLAS are consistent with the present result, 
although results of SAPHIR and CLAS are not consistent with each other\cite{bib:clas1,bib:sap1}.
The differential cross section decreases as $\sqrt{s}$  increases above $\sqrt{s}\sim$2.3~GeV in all the scattering angle 
bins.
The dashed lines in Fig.~3(e) and 3(f) represent the result of a power law fit for the Daresbury results\cite{bib:dar1}.
The smooth lines in Fig.~3 represent theoretical calculations which do not include the contribution of 
nucleon resonances\cite{bib:alex1}.
The parameters in these calculations were tuned to reproduce the Daresbury result which had a kinematic range of 
$E_{\gamma} = 2.8-4.8$~GeV and backward scattering.
It almost reproduces the Daresbury results except the very steep $s^{-8}$ dependence at $-0.2<u<-0.1$~GeV$^2$.
Since the differential cross section in the kinematic range of the Daresbury will be determined by 
the $u$-channel process,
the theoretical extrapolation to the LEPS energy range could be interpreted as the $u$-channel contribution.

The LEPS result shows that the present $\sqrt{s}$ dependence of $d\sigma/d\Omega$ is different from 
the theoretical extrapolation and the power law behavior for $\sqrt{s}\leq2.4$~GeV.
It implies that the present $\sqrt{s}$ dependence is difficult to explain by only the $u$-channel process, and that
there is a significant contribution from nucleon resonances via $s$-channel process in this energy range.
The theoretical curve becomes consistent with the LEPS and Daresbury 
result in $\sqrt{s}\geq$2.4~GeV.
The $\sqrt{s}$ dependence can be explained by only the $u$-channel process in $\sqrt{s}\geq$2.4~GeV.
To identify the possible nucleon resonances, we made a comparison between the present results and 
the theoretical extrapolation.
The present results show a broad excess of $d\sigma/d\Omega$  at 2$\leq\sqrt{s}\leq$2.4~GeV  compared with
the theoretical extrapolation at $-0.95<\cos\Theta_{C.M}^{\omega}<-0.80$.
The result of the Breit-Wigner fit for the excess depends on the $\omega$ scattering angle: its peak and width are 
2.25$\pm$0.04~GeV and 0.25$\sim$0.55~GeV, respectively.
The possible candidates are the G$_{17}(2190)$, H$_{19}(2220)$, and G$_{19}(2250)$, all of which 
have all 4-star states \cite{bib:pdg}.
Since all of these candidates have high angular momentum,
the angular distributions of these resonance decays could have rapid changes at backward angles.
It could account for the excess, which becomes smaller at the most backward scattering angle bin.
The coupling of the G$_{17}(2190)$ to $p\omega$ decay is supported by 
the PWA of $\gamma p \rightarrow p\omega$ at CLAS\cite{bib:pdg,bib:clas5}.
There are no reports for other candidates in the $\gamma p \rightarrow p\omega$ reaction.
The difference between the present data and the theoretical curve can be interpreted as 
an influence of the G$_{17}(2190)$.
However, this interpretation still does not explain the dip structure at 2$\leq\sqrt{s}\leq$2.1~GeV 
and $-1.00<\cos\Theta_{C.M}^{\omega}<-0.95$.
The present results may require more than just the G$_{17}(2190)$.
It is useful for the identification of nucleon resonances with high angler momentum to include
very backward angles in the PWA. 
It is worth mentioning that the very steep $s$ dependence of $\omega$ $d\sigma/du$  at 
$u\sim-0.15$~GeV$^2$, seen in the Daresbury data, are not observed in the LEPS energy range\cite{bib:dar1}.

\begin{figure}[thb]	
	\begin{center}	
	    \includegraphics[width=1\linewidth]{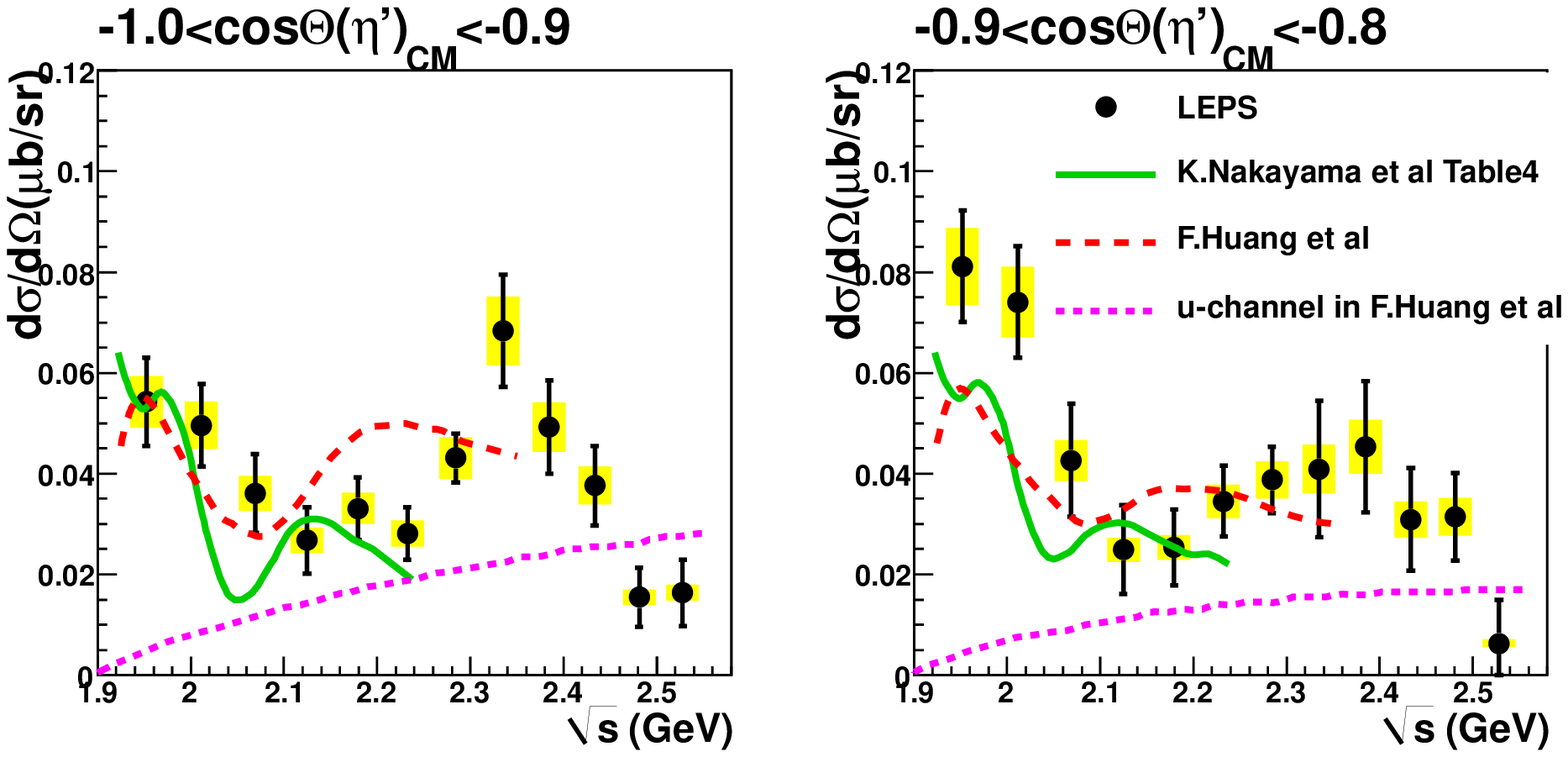}
	    \caption{(color online) Differential cross sections for $\eta'$ photoproduction 
	      a function of $\sqrt{s}$. 
	    Shaded bars represent systematic uncertainty. 
	    The black circles are the present results.
	    Smooth lines and dashed lines represent the theoretical calculation by 
	    K.~Nakayama\cite{bib:nh1} and F.~Huang\cite{bib:nh2}, respectively.
	    Dotted lines represent $u$-channel contribution in F.~Huang\cite{bib:nh2}.}
	  \end{center}
\end{figure}
Figure~4 shows the differential cross sections for $\eta'$ production as a function of $\sqrt{s}$. 
Each panel shows the result at the certain $\eta'$ scattering angle bin.
Theoretical calculations are also shown in Fig.~4. 
These calculations are tuned to reproduce the experimental results in a wide kinematic range 
except for the very forward and the backward angles\cite{bib:nh1,bib:nh2}.
The dotted lines in Fig.~4 show the $u$-channel contribution in the theoretical calculation by F.~Huang\cite{bib:nh2}.
It should be mentioned that the theoretical $u$-channel contribution at $\sqrt{s}>2.35$~GeV is just an extrapolation.
The $NN\eta'$  coupling constant $g_{NN\eta'}$ in this calculation was assumed to be 1.0.
While the best fit value for $g_{NN\eta'}$ depends on assumed resonances and parameters of the resonances, 
$g_{NN\eta'}$ was evaluated to be less than 2 by considering the extreme case\cite{bib:nh1}.
If we compare the present result with the $u$-channel contribution at $\sqrt{s}>2.4$~GeV,
$g_{NN\eta'}=1$ seems to be a reasonable assumption.
The dominant contribution at backward angles is from nucleon resonances 
at $\sqrt{s}<2.10$~GeV and the $\sqrt{s}\sim2.35$~GeV region according to 
the comparison between the LEPS result and the $u$-channel contribution.
The LEPS result of $\eta'$ shows an excess in $d\sigma/d\Omega$ 
at $\sqrt{s}\sim$2.35~GeV compared with the expected $u$-channel contribution.
This structure appears clearer in the most backward scattering angle bin, $-1.0<\cos\Theta_{C.M}^{\eta'}<-0.90$.
If the bump structure is also due to nucleon resonance, the corresponding resonance might have high angular momentum and 
a large branching ratio to $\eta'$.

Three nucleon resonances, the S$_{11}$(1535), P$_{13}$(1720), and D$_{15}$(2070), are suggested to have large branching 
ratio to $\eta$ and/or $\eta'$, but this depends on the model\cite{bib:ani1,bib:mai1}.
It is proposed that these resonances have the same quantum numbers except for the orbital angular momentum\cite{bib:ani1}.
According to this scenario, the fourth resonance with the strong coupling to $\eta$ and/or $\eta'$ may be 
a F$_{17}$ with a mass at around 2.3~GeV by the estimation from Regge trajectory.
There is a possibility that the observed bump structure is due to the F$_{17}$ resonances.
However, measurements including a wide kinematic range are necessary to explain the presence of the bump structure
at backward angles.

In summary, the $\omega$ and $\eta'$ photoproduction from protons at backward angle has been measured at $E_{\gamma}$ = $1.5-3.0$~GeV 
at the SPring-8/LEPS facility.
The $\omega$ and $\eta'$ mesons were identified by detecting forward scattered protons in the spectrometer and detecting pions from 
meson decay in the TPC surrounding the target. 
Background events for the $\omega$ and $\eta'$ signals were reduced substantially in comparison  with the previous
LEPS experiments by using a TPC.
The differential cross sections for $\omega$ mesons are larger than the expected $u$-channel contribution in the range $2.0\leq\sqrt{s}\leq2.4$~GeV.
Although a contribution from the G$_{17}(2190)$ resonance is a possible explanation for this excess in the $\omega$ differential cross sections,
a PWA including the present data is important to identify the possible resonances.
A bump structure in the $\sqrt{s}$ dependence of the differential cross sections for $\eta'$ mesons was observed at $\sqrt{s}\sim$2.35~GeV.
Measurements with a large acceptance is necessary to interpret the bump structure in the $\eta'$ differential cross sections.

We thank the staff at SPring-8 for providing excellent experimental conditions.
We thank A.~Sibirtsev and H.~Kamano for fruitful discussions. This work was supported in part by the Ministry of Education,
Science, Sports and Culture of Japan; the National Science Council of the Republic of China (Taiwan); 
the National Science Foundation (USA); 





\bibliographystyle{model1a-num-names}
\bibliography{<your-bib-database>}

\end{document}